\documentclass[conference]{IEEEtran}
\makeatletter
\def\ps@headings{%
\def\@oddhead{\mbox{}\scriptsize\rightmark \hfil \thepage}%
\def\@evenhead{\scriptsize\thepage \hfil \leftmark\mbox{}}%
\def\@oddfoot{}%
\def\@evenfoot{}}
\makeatother
\usepackage[utf8]{inputenc}

\usepackage{amsfonts,amsthm,amsmath,amssymb,verbatim,epsfig,psfrag,graphicx,graphics,caption,url}

\newtheorem*{theorem*}{Theorem}

\newtheorem*{lemma*}{Lemma}

\newtheorem*{proposition*}{Proposition}

\newtheorem{remark}{Remark}

\def\EE{{\mathbb E}}
\def\N{{\mathbb N}}
\def\R{{\mathbb R}}
\def\ind{\mathbf{1}}

\def\ln{\operatorname{ln}}
\def\rhoeff{{\rho_\text{eff}}}
\def\thetaeff{{\theta_\text{eff}}}

\title{Adaptive Replication in Distributed Content Delivery Networks}
\author{M. Leconte\\Technicolor - INRIA\\mathieu.leconte@inria.fr\and M. Lelarge\\INRIA - \'Ecole Normale Sup\'erieure\\marc.lelarge@ens.fr\and L. Massouli\'e\\Microsoft Research - INRIA Joint Center\\laurent.massoulie@inria.fr}

\begin{document}

\maketitle

\begin{abstract}
We address the problem of content replication in large distributed content delivery networks, composed of a data center assisted by many small servers with limited capabilities and located at the edge of the network. The objective is to optimize the placement of contents on the servers to offload as much as possible the data center. We model the system constituted by the small servers as a loss network, each loss corresponding to a request to the data center. Based on large system / storage behavior, we obtain an asymptotic formula for the optimal replication of contents and propose adaptive schemes related to those encountered in cache networks but reacting here to loss events, and faster algorithms generating virtual events at higher rate while keeping the same target replication. We show through simulations that our adaptive schemes outperform significantly standard replication strategies both in terms of loss rates and adaptation speed.
\end{abstract}

\section{Introduction}

The amount of multimedia traffic transfered over the Internet is steadily growing, mainly driven by video traffic. A large fraction of this traffic is already carried by dedicated content delivery networks (CDNs) and this trend is expected to continue \cite{cisco}. The largest CDNs are distributed, with a herd of storage points and data centers spread over many networks. With the cooperation of the Internet Service Providers, CDNs can deploy storage points anywhere in the network. However, these cooperations require a lot of negociations and are therefore only slowly getting established, and it also seems necessary to consider alternatives to a global cache-network planning. An easily deployable solution is to make use of the ressources (storage, upload bandwidth) already present at the edge of the network. Such an architecture offers a lot of potential for cost reduction and quality of service upgrade, but it also comes with new design issues \cite{DilleyMPPSW02} as the operation of a very distributed overlay of small caches with limited capabilities is harder than that of a more centralized system composed of only a few large data centers.

In this paper, we consider a hybrid CDN consisting of both a data center and an edge-assistance in the form of many small servers located at the edge of the network; we call such a system an edge-assisted CDN. The small servers are here to reduce the cost of operation of the CDN, while the data center helps provide reliability and quality of service guarantees. We stress the fact that the small servers have a limited storage and service capacity as well as limited coordination capabilities, which makes cheap devices eligible for that role. As their respective roles hint, we assume the service of a request by the data center is much more costly than by a small server, thus requests are served from the small servers in priority. In fact, we only send requests to the data center if no idle server storing the corresponding content is present. We assume the data center is dimensioned to absorb any such remaining flow of requests. We restrict our attention here to contents which require immediate service, which is why we do not allow requests to be delayed or queued. Stringent delay constraints are indeed often the case for video contents --for example in the context of a Video-on-Demand (VoD) service--, which represent already more than half of the Internet traffic.

In that context, we address the problem of determining which contents to store in the small servers. The goal is to find content placement strategies which offload as much as possible the data center. In this optic, it is obvious that popular contents should be more replicated, but there is still a lot of freedom in the replication strategy. In order to compare different policies, we model the system constituted by the small servers alone as a loss network, where each loss corresponds to a request to the data center. The relevant performance metric is then simply the expected number of losses induced by each policy.

We make three main contributions: first, using a mean-field heuristic for our system with a large number of servers with limited coordination, we compute accurately the loss rate of each content based on its replication; building on this analysis, we derive an optimized static replication policy and show that it outperforms significantly standard replications in the literature; finally, we propose adaptive algorithms attaining the optimal replication. Contrary to most caching schemes, our eviction rules are based on loss statistics instead of access statistics; in particular our algorithms do not know or learn the popularities of the contents and will adapt automatically. We also propose a simple mechanism relying heavily on our mean-field analysis that allows us to speed up significantly the rate of adaptation of our algorithms. At each step, extensive simulations support and illustrate our main points.

The rest of the paper is organized as follows: we review related work in Section~\ref{sec: related work} and describe more in detail our system model in Section~\ref{sec: model}. In Section~\ref{sec: analysis by approximation}, we analyse our model under a mean-field approximation and obtain an asymptotic expression for the optimal replication. Then, in Section~\ref{sec: replication policy}, leveraging the insights from the analysis we propose new adaptive schemes minimizing the average loss rate, as well as ways to further enhance their adaptation speed.

\section{Related Work}\label{sec: related work}

Our edge-assisted CDN model is related to a few other models. A major difference though is our focus on modeling the basic constraints of a realistic system, in particular regarding the limited capacity of servers (in terms of storage, service and coordination) and the way requests are handled, i.e. the \emph{matching} policy, which role is to determine which server an incoming request should be directed to. We indeed consider the most basic matching policy: an idle server storing the content is chosen uniformly at random; and no queueing of requests, no re-direction during service and no splitting are allowed. The reason for modeling such stringent constraints and refusing simplifying assumptions in this domain is that this should lead to fundamental intuitions and qualitative results that should apply to most practical systems.

Our edge-assisted CDN model is very similar to the distributed server system model of \cite{TanM11,LeconteLM12}; we merely use a different name to emphasize the presence in the core of the network of a data center, which makes it clear what performance metric to consider, while otherwise availability of contents for example may become a relevant metric. The edge-assisted CDN model is also related to peer-to-peer (P2P) VoD systems, which however have to deal with the behavior of peers, and to cache networks, where caches are deployed in a potentially complex overlay structure in which the flow of requests entering a cache is the ``miss'' flow of another one. Previous work on P2P VoD typically does not model all the fundamental constraints we mentioned, and the cache networks literature tends to completely obliterate the fact servers become unavailable while they are serving requests and to focus on alternative performance metrics such as search time or network distance to a copy of requested contents.

In distributed server networks, a maximum matching policy (with re-packing of the requests at any time but no splitting or queueing) has been considered in \cite{TanM11,LeconteLM12}. In this setting, in \cite{TanM11} it was shown that replicating contents proportionally to their popularity leads to full efficiency in large system and large storage limits, and in \cite{LeconteLM12} it was further proved that there is actually a wide range of policies having such an asymptotic property and an exact characterisation of the speed at which efficiency decreases for a given policy was obtained. However, that formula for the efficiency is too complicated to derive any practical scheme from it, and anyway maximum matching at any time is probably unrealistic for most practical systems.

Finding efficient content replications in cache networks is a very active area of research. The optimal replication is not yet understood, however adaptive strategies are studied; they create replicas for contents when incoming requests arrive and evict contents based on either random, least frequently used (LFU) or least recently used (LRU) rules, or a variation of those. There is an extensive literature on the subject and we only mention the most relevant papers to the present work. An approximation for LRU cache performance is proposed in \cite{che2002hierarchical} and further supported in \cite{fricker2012versatile}; it allows \cite{fricker2012impact} to study cache networks with a two-levels hierarchy of caches under a mix of traffic with various types of contents. Based on the differences in their respective popularity distributions, \cite{fricker2012impact} advocates that VoD contents are cached very close to the network edge while user-generated contents, web and file sharing are only stored in the network core. In a loose sense, we also study such type of two-layer systems, with the small servers at the edge of the network and the data center in the core, and address the question of determining whether to store contents at the edge (and in what proportion) or to rely on the core to serve them.

Concerning P2P VoD models, content pre-fetching strategies have been investigated in many papers (e.g. \cite{TanM11,ZhouFC13,ZhouFC12,CiulloMGLT12}), where optimal replication strategies and/or adaptive schemes were derived. They work under a maximum matching assumption and most of them assume infinite divisibility of the requests, i.e. that a request can be served in parallel by all the servers storing the corresponding content; we want to take into account more realistic constraints in that respect.

One can also explore the direction of optimizing distributed systems with a special care for the geographical aspects as in \cite{jiang2012orchestrating,rochman2013resource,rochman2012max}. These papers solve content placement and matching problems between many regions, while not modeling in a detailed way what happens inside a region.

Finally, most of the work in these related domains focuses on different performance metrics: in hierarchical cache networks, \cite{laoutaris2005optimization} addresses the problem of joint dimensioning of caches and content placement in order to reduce overall bandwidth consumed; \cite{cohen2002replication} optimizes replication in order to minimize search time and devises elegant adaptive strategies towards that end. Surprisingly, for various alternative objectives and network models, the proportional (to popularity) replication exhibits good properties: it minimizes the download time in a fully connected network with infinite divisibility of requests and convex server efficiency \cite{tewari2005fairness}, as well as the average distance to the closest cache holding a copy of the requested content in a random network with exponential expansion \cite{tewari2006proportional}; and recall that \cite{TanM11} also advocated using this replication policy. Therefore, due to its ubiquity in the literature, proportional replication is the natural and most relevant scheme to which we should compare any proposed policy; we will obtain substantial improvements over it in our model.

\section{Edge-Assisted CDN Model}\label{sec: model}

In this section, we describe in detail our edge-assisted CDN model.

As already mentioned, the basic components of an edge-assisted CDN are a data center and a large number $m$ of small servers. The CDN offers access to a catalog of $n$ contents, which all have the same size for simplicity (otherwise, they could be fragmented into constant-size segments). The data center stores the entire catalog of contents and can serve all the requests directed towards it, whereas each small server can only store a fixed number $d$ of contents and can provide service for at most one request at a time. We can represent the information of which server stores which content in a bipartite graph $G=(S\cup C,E)$, where $S$ is the set of servers, $C$ the set of contents, and there is an edge in $E$ between a content $c$ and a server $s$ if $s$ stores a copy of $c$; an edge therefore indicates that a server is able to serve requests for the content with which it is linked. Given numbers of replicas $(D_c)_{c\in C}$ for the contents, we assume that the graph $G$ is a uniform random bipartite graph with fixed degrees $d$ on the servers side and fixed degree sequence $(D_c)_{c\in C}$ on the contents side. This models the lack of coordination between servers and would result from them doing independent caching choices. We do not allow requests to be queued, delayed or re-allocated during service, and as a result a server which begins serving a request will then remain unavailable until it completes its service. Splitting of requests is not considered either, in the sense that only one server is providing service for a particular request. With these constraints, at any time, the subset of edges of $G$ over which some service is being provided form a generalized matching of the graph $G$ (the analysis in \cite{LeconteLM12,LeconteLM13} relies heavily on this).

The contents of the catalog may have different popularities, leading to different requests arrival rates $\lambda_c$. We let $\overline\lambda$ be the average request rate, i.e. $\overline\lambda=\frac{1}{n}\sum_c\lambda_c$. According to the independent reference model (IRM), the requests for the various contents arrive as independent Poisson processes with rates $\lambda_c$. The service times of all the requests are independent Exponential random variables with mean $1$, which is consistent with all the servers having the same service capacity (upload bandwidth) and all the contents having the same size. We let $\rho$ denote the expected load of the system, i.e. $\rho=\frac{n\overline\lambda}{m}$. The distribution of popularities of the contents is left arbitrary, although in practice Zipf distributions are often encountered (see \cite{fricker2012impact} for a page-long survey of many studies) and it seems therefore mandatory to evaluate the replication schemes proposed under a Zipf popularity distribution.

When a request arrives, we first look for an idle server storing the corresponding content; if none is found then the request is sent to the data center to be served at a higher cost. As in the end all the requests are served without delay, be it by the data center or by a small server, the performance of the system is mainly described by the cost associated with the service. This cost is mostly tied to the number of requests served by the data center, therefore the most relevant performance metric here is the fraction of the load left-over to the data center. Then, it makes sense to view the requests that the small servers cannot handle as ``lost''. In fact, the system consisting of the small servers alone with fixed caches is a loss network in the sense of \cite{kelly1991loss}. This implies that the system corresponds to a reversible stochastic process, with a product form stationary distribution $\pi$. However, the exact expression of that stationary distribution is too complex to be exploited in our case (as opposed to \cite{TanM11,LeconteLM12,LeconteLM13} where the maximum matching assumption yields a much simpler expression). We call $\gamma_c$ the rate at which requests for content $c$ are lost, and $\overline\gamma$ the average loss rate among contents, i.e. $\overline\gamma=\frac{1}{n}\sum_c\gamma_c$. The main goal is to make $\overline\gamma$ as small as possible. We refer to the fraction of lost requests $\overline\gamma/\overline\lambda$ as the inefficiency of the system.

Finally, as in many real cases CDNs are quite large, with often thousands of servers and similar numbers of contents, it seems reasonable to pay a particular attention to the asymptotics of the performance of the system as $n,m\to\infty$. To keep things simple, we can assume that the number of servers $m$ and the number of contents $n$ grow to infinity together, i.e. $n=\beta m$ for some fixed $\beta$. In addition, as storage is cheap nowadays compared to access bandwidth, it also makes sense to focus on a regime with $d$ large (but still small compared to $n$). In these regimes, the inefficiency will tend to $0$ as the system size increases under most reasonable replications (as was shown in \cite{LeconteLM12} under the maximum matching assumption). However, as the cost of operation is only tied to losses, further optimizing an already small inefficency is still important and can lead to order improvements on the overall system cost.

\section{An Approximation for the Loss Rate of a Content}\label{sec: analysis by approximation}

In this section, we propose an approximation to understand in a precise manner the relation between any fixed replication of contents and the loss rates in the system. This analytical step has many advantages: it allows us to formally demonstrate that to optimize the system one needs to make the loss rates equal for all the contents; as a consequence we obtain an explicit expression for the optimal replication (Subsection~\ref{sec: optimized replication}); finally, in Subsection~\ref{sec: virtual losses}, we will leverage our analytical expression for the full distribution of the number of available replicas of a content to propose a mechanism enhancing the speed of adaptive algorithms. We validate our approximation and show that our optimized replication strategy largely outperforms proportional replication through extensive simulations.

\subsection{One-Dimensional Markov Chain Approximation for the Number of Available Replicas}

For a given fixed constitution of the caches (i.e. a fixed graph $G$), the system as described in the previous section is Markovian, the minimum state space indicating which servers are busy (it is not necessarily to remember which content they serve). We want to understand how the loss rate $\gamma_c$ for a particular content $c$ relates to the graph $G$, but using too detailed information about $G$, such as the exact constitution of the caches containing $c$, would have the drawback that it would not lead to a simple analysis when considering adaptive replication policies (as the graph $G$ would then keep changing). Therefore, we need to obtain a simple enough but still accurate approximate model tying $\gamma_c$ to $G$.

The expected loss rate $\gamma_c$ of content $c$ is equal to its requests arrival rate $\lambda_c$ multiplied by the steady state probability that $c$ has no replicas available. Let $D_c$ be the total number of replicas of content $c$ and $Z_c$ be its number of \emph{available} replicas, i.e. those stored on a currently idle server. We thus want to compute $\pi(Z_c=0)$ to get access to $\gamma_c$. However, the Markov chain describing the system is too complicated to be able to say much on its steady state. In order to simplify the system, one can remark that in a large such system the state of any fixed number of servers (i.e. their current caches and whether they are busy or idle) are only weakly dependent, and similarly the number of available replicas $Z_c$ of any fixed number of contents are only weakly dependent. Therefore, it is natural to approximate the system by decoupling the different contents and the different servers (similar phenomenon are explained rigorously in \cite{sznitman1991topics}). In other words, this amounts to forgetting the exact constitution of the caches; as a result, the correlation between contents which are stored together is lost. Then, the evolution of $Z_c$ becomes a one dimensional Markov chain, independent of the values of $Z_{c'}$ for other contents, and we can try to compute its steady-state distribution.

For any $z<D_c$, the rate of transition from $Z_c=z$ to $z+1$ is always $D_c-z$: it is the rate at which one of the $D_c-z$ occupied servers storing $c$ completes its current service; we do not need to distinguish whether a server is actually serving a request for $c$ or for another content $c'$ as we assume the service times are independent and identically distributed across contents. For any $z>0$, the transitions from $Z_c=z$ to $z-1$ are more complicated. They happen in the two following cases:
\begin{itemize}
\item either a request arrives for content $c$ and as $Z_c=z>0$ it is assigned an idle servers storing $c$;
\item or a request arrives for another content $c'$ and it is served by a server which also stores content $c$.
\end{itemize}
The first event occurs at rate $\lambda_c$ and the second one at expected rate
\begin{equation*}
\sum_{c'\neq c}\lambda_{c'}\EE\left[\frac{|\{s\in S:\:s\text{ is idle and }c,c'\in s\}|}{|\{s\in S:\:s\text{ is idle and }c'\in s\}|}\right],
\end{equation*}
where $c\in s$ indicates that the content $c$ is stored on the server $s$. At any time and for any $c'$, $|\{s\in S:\:s\text{ is idle and }c'\in s\}|$ is equal to $Z_{c'}$. The term $|\{s\in S:\:s\text{ is idle and }c,c'\in s\}|$ is equal in expectation to the number of idle servers storing $c'$ (i.e. $Z_c'$) times the probability that such a server also stores $c$. As we forget about the correlations in the caching, this last probability is approximated as the probability to pick one of the $d-1$ remaining memory slots in an idle servers storing $c'$ when we dispatch at random the $Z_c$ available replicas of content $c$ between all the remaining slots in idle servers. Thus,
\begin{equation*}
\EE\left[|\{s\in S:\:s\text{ is idle and }c,c'\in s\}|\right]\approx\frac{Z_{c'}(d-1)}{(1-\rhoeff)md}Z_c,
\end{equation*}
where $\rhoeff=\rho(1-\overline\gamma/\overline\lambda)$ is the average load effectively absorbed by the system, so that the total number of memory slots on the idle servers is $(1-\rhoeff)md$. We also neglected the $Z_{c'}$ memory slots occupied by $c'$ in these idle servers when computing the total number of idle slots $(1-\rhoeff)md$. We obtain the following approximation for the expected rate at which the second event occurs:
\begin{equation*}
Z_c\sum_{c'\neq c}\lambda_{c'}\frac{d-1}{(1-\rhoeff)md}\approx Z_c\frac{\rhoeff}{1-\rhoeff}\frac{d-1}{d}=Z_c\thetaeff,
\end{equation*}
where we neglected the rate $\lambda_c$ at which requests arrive for content $c$ compared to the aggregate arrival rate of requests, and we let $\thetaeff=\frac{\rhoeff}{1-\rhoeff}\frac{d-1}{d}$. Note that the interesting regime, with reasonably high effective load $\rhoeff$, corresponds to large values of $\thetaeff$, as $\rhoeff\to1$ implies $\thetaeff\to\infty$. The Markov chain obtained satisfies the local balance equations: for $z<D_c$,
\begin{equation*}
(D_c-z)\pi(Z_c=z)=(\lambda_c+(z+1)\thetaeff)\pi(Z_c=z+1),
\end{equation*}
This yields the following steady-state probability:
\begin{equation}\label{eqn: approximation Z_c}
\pi(Z_c=z)=\pi(Z_c=D_c)\theta_\text{eff}^{D_c-z}\binom{D_c+\lambda_c/\thetaeff}{D_c-z},
\end{equation}
where the binomial coefficient does not really have a combinatorial interpretation as one of its arguments is not an integer, but should only be understood as $\binom{k+x}{l}=\frac{1}{l!}\prod_{i=k-l+1}^k(i+x)$, for $k,l\in\N$ and $x\in\R_+$.

We now have an approximation for the full distribution of the number $Z_c$ of available replicas of content $c$, which yields an approximation for the loss rate $\gamma_c=\lambda_c\pi(Z_c=0)$. We can also compute the mode $\widehat z_c$ of this distribution: $\widehat z_c\approx\frac{D_c-\lambda_c}{1+\thetaeff}$, which can be used as an approximation for the expectation $\EE[Z_c]$ (in fact, simulations show the two are nearly indistinguishable). We further simplify the expression obtained by providing a good approximation for the normalization factor $\pi(Z_c=D_c)$ in Equation~(\ref{eqn: approximation Z_c}):
\begin{equation*}\label{eqn: normalizing constant}
\pi(Z_c=D_c)=\left(\sum_{x=0}^{D_c}\theta_\text{eff}^x\binom{D_c+\lambda_c/\thetaeff}{x}\right)^{-1}.
\end{equation*}
To that end, we use the fact that we aim at small loss rates, and thus most of the time at small probabilities of unavailability. Therefore, the bulk of the mass of the distribution of $Z_c$ should be away from $0$, which means that the terms for $x$ close to $D_c$ should be fairly small in the previous expression. We thus extend artificially the range of the summation in this expression and approximate $\pi(Z_c=D_c)$ as follows:
\begin{eqnarray*}
\pi(Z_c=D_c)^{-1}&\approx&\sum_{x=0}^{D_c+\lfloor\lambda_c/\thetaeff\rfloor}\theta_\text{eff}^x\binom{D_c+\lambda_c/\thetaeff}{x}\\
&\approx&(1+\thetaeff)^{D_c+\lambda_c/\thetaeff}.
\end{eqnarray*}
We obtain the following approximation for $\pi(Z_c=0)$:
\begin{equation*}
\pi(Z_c=0)\approx\left(\frac{\thetaeff}{1+\thetaeff}\right)^{D_c}(1+\thetaeff)^{-\frac{\lambda_c}{\thetaeff}}\binom{D_c+\lambda_c/\thetaeff}{D_c}.
\end{equation*}
Finally, as we are interested in large systems and large storage / replication, we can leverage the following asymptotic behavior:
\begin{equation*}
\binom{k+x}{k}\approx\frac{e^{x(H_k-\text{C}_\text{Euler})}}{\Gamma(x+1)},
\end{equation*}
for $k\in\N$ large enough and where $\Gamma$ is the gamma function: $\Gamma(1+x)=\int_0^{+\infty}t^x e^{-t}\text dt$; $H_k$ is the $k$-th harmonic number: $H_k=\sum_{i=1}^k\frac{1}{i}$; and $\text{C}_\text{Euler}$ is the Euler-Mascheroni constant: $\text{C}_\text{Euler}=\lim_{k\to\infty}(H_k-\ln k)\approx0.57721$. For large number of replicas $D_c$, we thus obtain the approximation:
\begin{equation*}
\pi(Z_c=0)\approx\text C(\lambda_c)\left(1+\frac{1}{\thetaeff}\right)^{-D_c}D_c^\frac{\lambda_c}{\thetaeff},\label{eqn: unavailability probability}
\end{equation*}
where we let $\text C(\lambda_c)=\frac{e^{-\text{C}_\text{Euler}\lambda_c/\thetaeff}}{(1+\thetaeff)^\frac{\lambda_c}{\thetaeff}\Gamma(1+\frac{\lambda_c}{\thetaeff})}$ and the term $D_c^\frac{\lambda_c}{\thetaeff}$ is an approximation of $e^{\frac{\lambda_c}{\thetaeff}H_{D_c}}$. The approximation for $\pi(Z_c=0)$ immediately yields an approximation for the loss rate $\gamma_c$:
\begin{equation}\label{eqn: loss rate}
\gamma_c=\lambda_c\pi(Z_c=0)\approx\lambda_c\text C(\lambda_c)\left(1+\frac{1}{\thetaeff}\right)^{-D_c}D_c^\frac{\lambda_c}{\thetaeff}.
\end{equation}
Note that the expression of $\thetaeff$ involves the average effective load $\rhoeff$, which itself depends on the average loss rate $\overline\gamma$. We thus have a fixed point equation in $\overline\gamma$ (just as in \cite{che2002hierarchical,fricker2012versatile}), which we can easily solve numerically. Indeed, the output value $\overline\gamma$ of our approximation is a decreasing function of the input value $\overline\gamma$ used in $\rhoeff$, which implies simple iterative methods converge exponentially fast to a fixed-point value for $\overline\gamma$. 

\subsection{Approximation-Based Optimized Replication}\label{sec: optimized replication}

In this subsection, we exploit the approximation obtained in Equation~(\ref{eqn: loss rate}) to understand which replication strategy minimizes the total loss rate in the system. In other words, we approximately solve the optimization problem:
\begin{equation}\label{eqn: optimization problem}
\begin{array}{clc}
\min&\overline\gamma&\text{s.t.}
\begin{array}{l}
\sum_c D_c\leq md\\
D_c\in\N,\:\forall c
\end{array}
\end{array}
\end{equation}
Note that the approximation from Equation~(\ref{eqn: loss rate}) is consistent with the intuition that $\gamma_c$ is a decreasing, convex function of $D_c$. Indeed, letting $x=\frac{\thetaeff}{1+\thetaeff}e^\frac{\lambda_c}{\thetaeff(D_c+1)}\leq1$ since we have $D_c+1\geq\frac{\lambda_c}{\thetaeff\ln\left(1+\frac{1}{\thetaeff}\right)}$ in the regime of interest, we compute
\begin{equation*}
\begin{array}{r}
\gamma_c(D_c+1)-\gamma_c(D_c)=\Delta\gamma_c(D_c)=-\gamma_c(D_c)\left(1-x\right)\leq0,\\
\Delta\gamma_c(D_c-1)=-\gamma_c(D_c)\left(\frac{1}{x}-1\right)\leq0.
\end{array}
\end{equation*}
The loss rate $\gamma_c$ is a convex function of $D_c$ as shown by
\begin{equation*}
\Delta\gamma_c(D_c)-\Delta\gamma_c(D_c-1)\geq\gamma_c(D_c)\left(x+\frac{1}{x}-2\right)\geq0.
\end{equation*}
As a consequence, the optimization problem~(\ref{eqn: optimization problem}) is approximately convex, and we thus obtain an approximate solution by making the first derivatives of the loss rates $\Delta\gamma_c$ equal. Keeping only the dominant orders in $D_c$, we have
\begin{equation}\label{eqn: first derivative}
\Delta\gamma_c(D_c)=-\frac{\gamma_c}{1+\thetaeff}\left(1-\frac{\lambda_c}{D_c}\right).
\end{equation}
In the first order, equalizing the first derivatives in~(\ref{eqn: first derivative}) using the expression in~(\ref{eqn: loss rate}) leads to equalizing the number of replicas for every content, i.e. setting $D_c=\overline D+o(\overline D))$ where $\overline D=\frac{md}{n}$ is the average number of replicas of contents. Going after the second order term, we get
\begin{equation}\label{eqn: optimal replication}
D_c=\overline D+(\lambda_c-\overline\lambda)\frac{\ln\overline D}{\thetaeff\ln\left(1+\frac{1}{\thetaeff}\right)}+o(\ln\overline D).
\end{equation}
We therefore obtain that the asymptotically optimal replication is uniform with an adjustment due to popularity of the order of the log of the average number of replicas. This agrees with the results in \cite{LeconteLM12} and goes beyond.
Finally, inserting back this expression for $D_c$ into Equation~(\ref{eqn: loss rate}) yields
\begin{equation}\label{eqn: optimal gamma}
\gamma_c=\left(1+\frac{1}{\thetaeff}\right)^{-\overline D}\overline D^{\frac{\overline\lambda}{\thetaeff}(1+o(1))},
\end{equation}
which shows that the average loss rate $\overline\gamma$ under optimized replication behaves as the loss rate of an imaginary average content (one with popularity $\overline\lambda$ and replication $\overline D$).

\subsection{Accuracy of the Approximation and Performance of the Optimized Replication}\label{sec: simulation approximation}

In the process of approximating the loss rate $\gamma_c$ of a content $c$, we performed many approximations based on asymptotic behaviors for large systems and large storage / replication. It is therefore necessary to check whether the formula of Equation~(\ref{eqn: loss rate}) is not too far off, which we do in this section via simulation. The systems we simulate are of quite reasonable size (with a few hundred contents and servers). However, the accuracy of the approximation should only improve as the system grows. We use two scenarios for the popularity distribution of the contents: a class model (similar to that of \cite{LeconteLM12}), which allows us to compare between many contents with similar characteristics, and a Zipf popularity model, which is deemed more realistic. We evaluate the accuracy of the approximation using proportional replication (other replications were also simulated and yield similar results). To compute the approximate expressions, we solve numerically the fixed point equation in $\overline\gamma$. We simulate systems under a reasonably high load of $0.9$. As often, it is mainly interesting to know how the system behaves when the load is high, as otherwise its performance is almost always good. However, as requests arrive and are served stochastically, if the average load were too close to $1$ then the instantaneous load would exceed $1$ quite often, which would automatically induce massive losses and mask the influence of the replication. In fact, it is easy to see that we need to work with systems with a number of servers $m$ large compared to $\frac{\rho}{(1-\rho)^2}$ in order to mitigate this effect.

\begin{table}
\begin{center}
\begin{tabular}{| ccccccc |}
\hline
Properties & \multicolumn{2}{|c|}{class 1} & \multicolumn{2}{|c|}{class 2} & \multicolumn{2}{|c|}{class 3} \\
\hline 
number of contents $n_i$ & \multicolumn{2}{|c|}{200}  & \multicolumn{2}{|c|}{400}  & \multicolumn{2}{|c|}{400}  \\
\hline
popularity $\lambda_i$ & \multicolumn{2}{|c|}{9}  &  \multicolumn{2}{|c|}{3}  & \multicolumn{2}{|c|}{1}  \\
\hline
number of replicas $D_i$ & \multicolumn{2}{|c|}{200}  &  \multicolumn{2}{|c|}{67}  & \multicolumn{2}{|c|}{23}  \\
\hline
\hline
simulation data / approximation& \multicolumn{2}{|c|}{class 1} & \multicolumn{2}{|c|}{class 2} & \multicolumn{2}{|c|}{class 3} \\
\hline 
nb. of available replicas $\EE[Z_i]$&\multicolumn{2}{|c|}{21.7 / 21.6}&\multicolumn{2}{|c|}{7.28 / 7.25}&\multicolumn{2}{|c|}{2.51 / 2.50}\\
\hline
$10^3\times$ loss rates $\gamma_i$ &\multicolumn{2}{|c|}{0 / $10^{-5}$}&\multicolumn{2}{|c|}{3.31 / 2.36}&\multicolumn{2}{|c|}{79.4 / 76.3}\\
\hline
\hline
simulation data / approximation&\multicolumn{3}{|c|}{ Zipf, $\alpha=0.8$ }&\multicolumn{3}{|c|}{ Zipf, $\alpha=1.2$ }\\
\hline
proportional: $10^3\times$ inefficiency &\multicolumn{3}{|c|}{5.69 / 5.46}&\multicolumn{3}{|c|}{11.8 / 11.6}\\
\hline
\end{tabular}
\caption{Properties of the content classes, and numerical results on the accuracy of the approximation.}
\label{table: class properties}
\end{center}
\end{table}

The setup with the class model is the following: there are $n=1000$ contents divided into $K=3$ classes; the characteristics of the classes are given in the first part of Table~\ref{table: class properties}. The popularities in Table~\ref{table: class properties} together with $n=1000$ and $\rho=0.9$ result in $m=3800$ servers. Each server can store $d=20$ contents, which is $5\%$ of the catalog of contents. We let the system run for $10^4$ units of time, i.e. contents of class 3 should have received close to $10^4$ requests each.

\begin{figure}
\centering
\[
\begin{array}{ccc}
\includegraphics[width=1.6in,keepaspectratio]{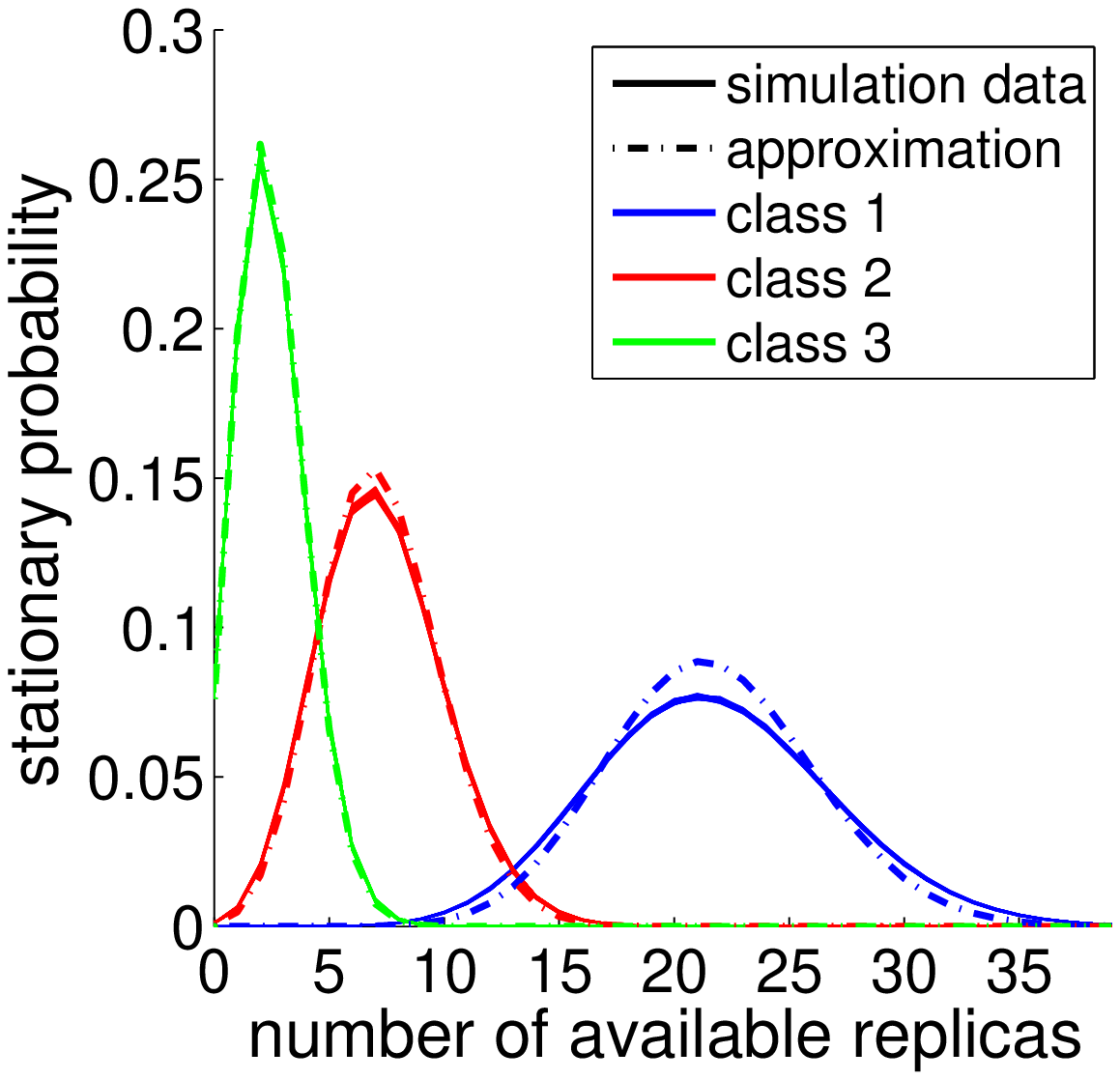}&
\includegraphics[width=1.7in,keepaspectratio]{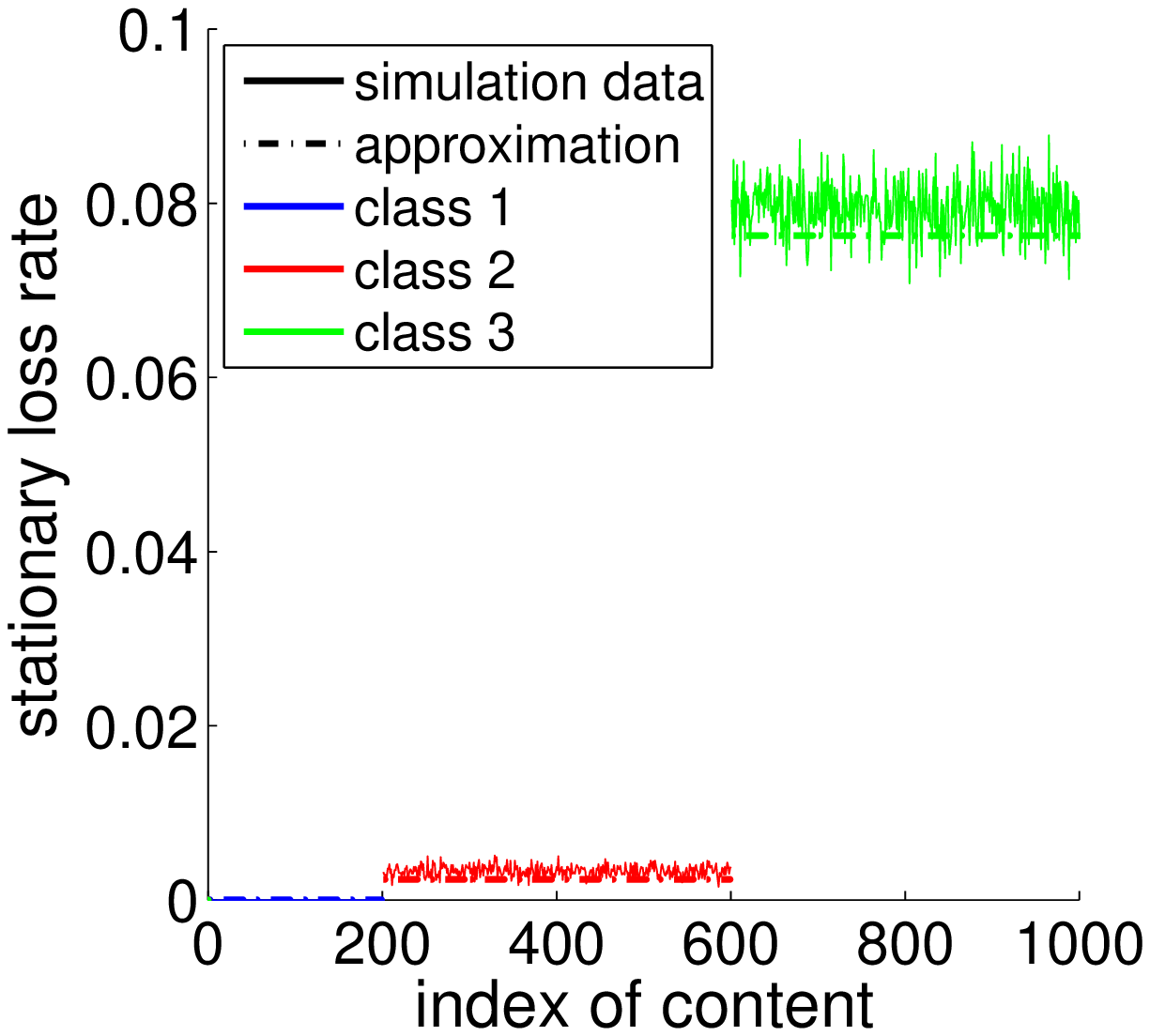}
\end{array}
\]
\caption{Class model: distribution of number of available replicas and loss rates Vs approximation.}
\label{fig: simulation Vs approximation, class model}
\end{figure}

%

Figure~\ref{fig: simulation Vs approximation, class model} and the second part of Table~\ref{table: class properties} show the results for the class model: the left part of Figure~\ref{fig: simulation Vs approximation, class model} shows the distributions of the numbers of available replicas for all the contents against the approximation from Equation~(\ref{eqn: approximation Z_c}) for each class; the right part shows the loss rates for all the contents against the approximation from Equation~(\ref{eqn: loss rate}) for each class. Although it is not apparent at first sight, the left part of Figure~\ref{fig: simulation Vs approximation, class model} actually displays a plot for each of the $1000$ contents, but the graphs for contents of the same class overlap almost perfectly, which supports our approximation hypothesis that the stationary distribution of the number of available replicas is not very dependent on the specific servers on which a content is cached or on the other contents with which it is cached. This behavior is also apparent on the right part of Figure~\ref{fig: simulation Vs approximation, class model}, as the loss rates for the contents of the same class are quite similar. From Figure~\ref{fig: simulation Vs approximation, class model} and Table~\ref{table: class properties}, it appears that the approximations from Equations~(\ref{eqn: approximation Z_c}) and~(\ref{eqn: loss rate}) are quite accurate, with for example a relative error of around $5\%$ in the inefficiency of the system ($9.68\times 10^{-3}$ Vs $9.20\times 10^{-3}$). We consider such an accuracy is quite good given that some of the approximations done are based on a large storage / replication asymptotic, while the simulation setup is with a storage capacity of $d=20$ contents only and contents of class $3$ (responsible for the bulk of losses) have only $23$ replicas each.

We now turn to the Zipf popularity model. In this model, the contents are ranked from the most popular to the least; for a given exponent parameter $\alpha$, the popularity of the content of rank $i$ is given by $\lambda_i=\frac{i^{-\alpha}}{\sum_{j\leq n}j^{-\alpha}}\overline\lambda$. We use two different values for the Zipf exponent $\alpha$, $0.8$ and $1.2$, as proposed in \cite{fricker2012impact}. The exponent $0.8$ is meant to represent correctly the popularity distribution for web, file sharing and user generated contents, while the exponent $1.2$ is more fit for video-on-demand, which has a more accentuated popularity distribution. We simulate networks of $n=200$ contents and $m\approx2000$ servers of storage capacity $d=10$ under a load $\rho=0.9$. This yields an average content popularity $\overline\lambda=\rho\frac{m}{n}\approx9$. Note that under proportional replication the numbers of replicas of the most popular contents are actually larger than the number of servers $m$ for $\alpha=1.2$; we thus enforce that no content is replicated in more than $95\%$ of the servers. As expected and confirmed by the simulations, this is anyway benefical to the system. Each setup is simulated for at least $10^4$ units of time.

\begin{figure}
\centering
\[
\begin{array}{ccc}
\includegraphics[width=1.6in,keepaspectratio]{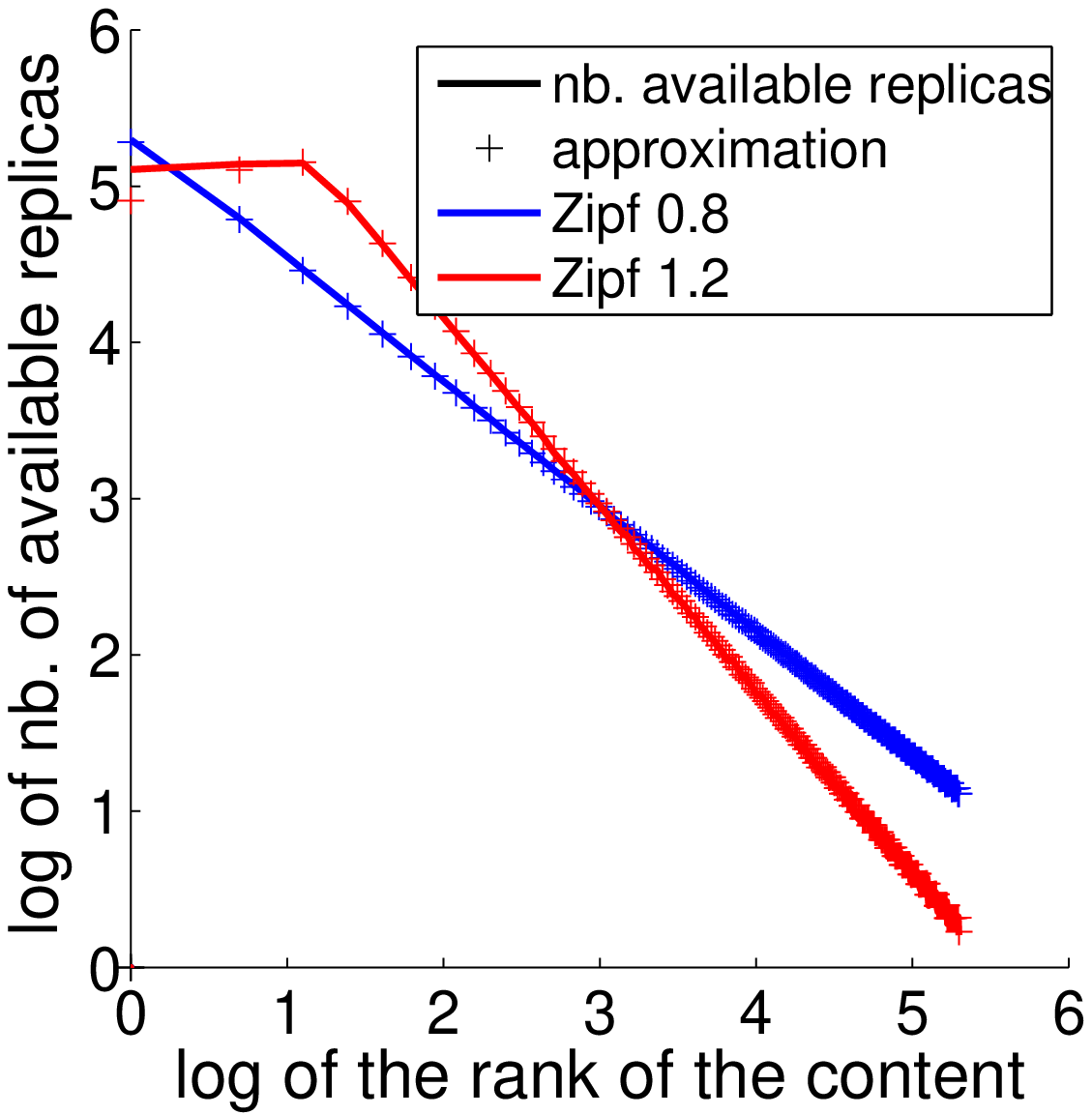}&
\includegraphics[width=1.7in,keepaspectratio]{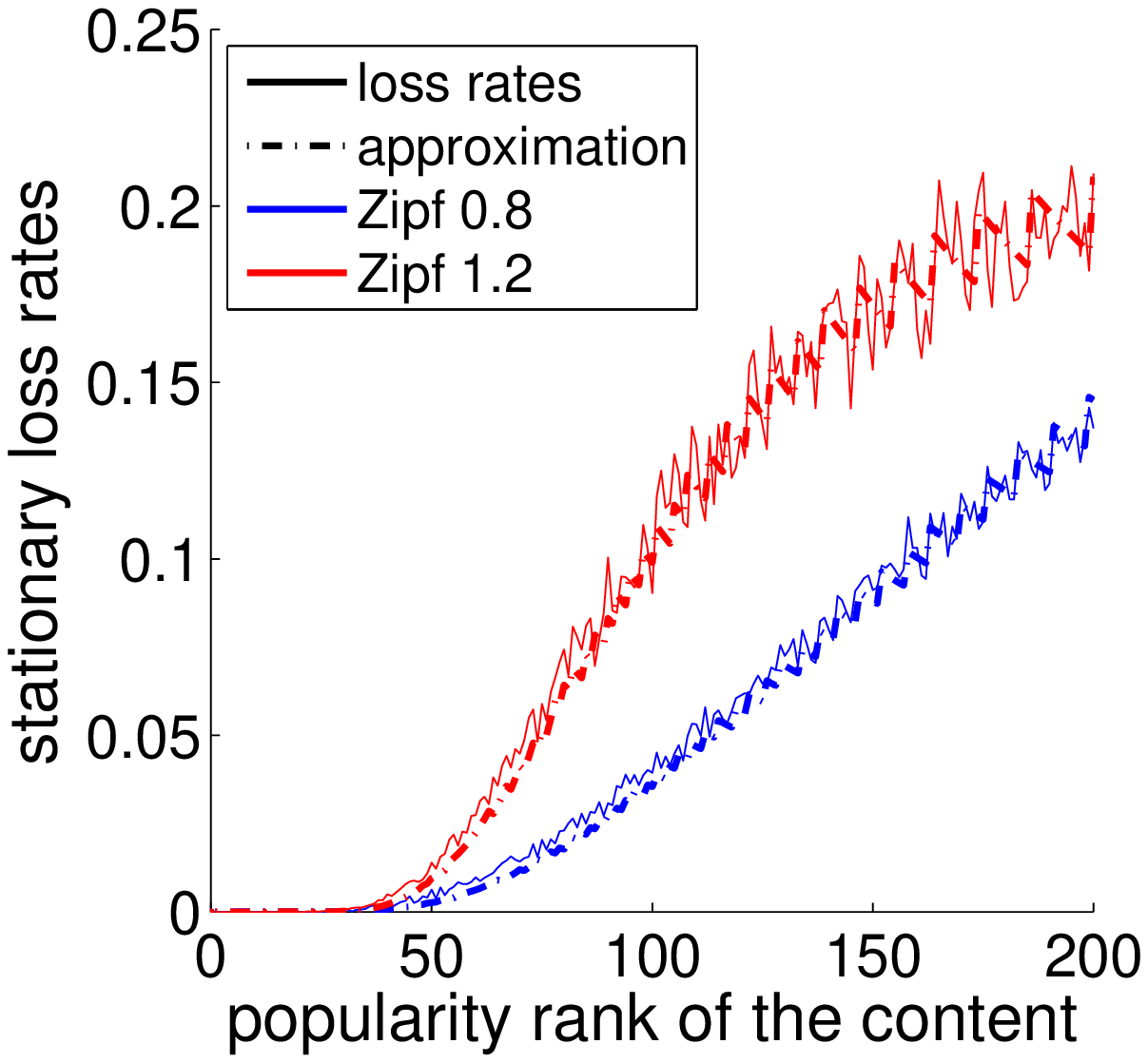}
\end{array}
\]
\caption{Zipf model: expected number of available replicas and loss rates Vs approximation.}
\label{fig: simulation Vs approximation, Zipf model}
\end{figure}

%

We show the results for the Zipf model with both exponent values in Figure~\ref{fig: simulation Vs approximation, Zipf model} and in the third part of Table~\ref{table: class properties}: the left part of Figure~\ref{fig: simulation Vs approximation, Zipf model} shows the expected number of available replicas for all the contents against the approximation from Equation~(\ref{eqn: approximation Z_c}); the right part shows the loss rates for all the contents against the approximation from Equation~(\ref{eqn: loss rate}). Again, the results from both Figure~\ref{fig: simulation Vs approximation, Zipf model} and Table~\ref{table: class properties} confirm the accuracy of the approximations.

\begin{figure}
\centering
\includegraphics[width=2.3in,keepaspectratio]{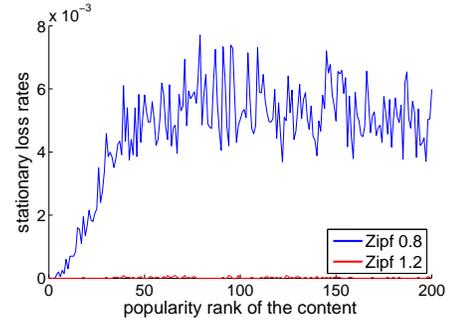}
\caption{Zipf model: loss rates under optimized replication.}
\label{fig: optimized replication}
\end{figure}

We now turn to evaluating the performance of the optimized replication from Equation~(\ref{eqn: optimal replication}). Figure~\ref{fig: optimized replication} shows the resulting loss rates under the optimized replication, with both Zipf exponents. This figure is to be compared with the right part of Figure~\ref{fig: simulation Vs approximation, Zipf model}, which showed the same results for proportional replication. It is clear that the optimized replication succeeds at reducing the overall inefficiency $\overline\gamma/\overline\lambda$ compared to proportional replication (from $5.7\times 10^{-3}$ to $5.1\times 10^{-4}$ for $\alpha=0.8$, and from $1.2\times 10^{-2}$ to $1.6\times 10^{-6}$ for $\alpha=1.2$). Note that in the case of Zipf exponent $\alpha=1.2$ the popularity distribution is more ``cacheable'' as it is more accentuated, and the optimized replication achieves extremely small loss rates. However, the loss rates for all the contents are not trully equalized, as popular contents are still too much favored (as can be seen for Zipf exponent $\alpha=0.8$). This may be because the expression for optimized replication is asymptotic in the system size and storage capacity. Hence, there is still room for additional finite-size corrections to the optimized replication.

As an outcome of this section, we conclude that the approximations proposed are accurate, even at reasonable system size. In addition, the optimized scheme largely outperforms proportional replication. In the next section, we derive adaptive schemes equalizing the loss rates of all the contents and achieving similar performances as the optimized replication.

\section{Designing Adaptive Replication Schemes}\label{sec: replication policy}

In a practical system, we would want to adapt the replication in an online fashion as much as possible, as it provides more reactivity to variations in the popularity of contents. Such variations are naturally expected for reasons such as the introduction of new contents in the catalog or a loss of interest for old contents. Thus, blindly enforcing the replication of Equation~(\ref{eqn: optimal replication}) is not always desirable in practice. Instead, we can extract general principles from the analysis performed in Section~\ref{sec: analysis by approximation} to guide the design of adaptive algorithms.

In this section, we first show how basic adaptive rules from cache networks can be translated into our context based on the insights from Section~\ref{sec: analysis by approximation} to yield adaptive algorithms minimizing the overall loss rate. Then, we show how the more detailed information contained in Equation~(\ref{eqn: approximation Z_c}) allows the design of faster schemes attaining the same target replication. Finally, we validate the path followed in this paper by evaluating through simulations the adaptive schemes proposed.

\subsection{Simple Adaptive Algorithms for Minimizing Losses}
The analysis in Section~\ref{sec: optimized replication} shows that the problem of minimizing the average loss rate $\overline\gamma$ is approximately a convex problem, and that therefore one should aim at equalizing the derivatives $\Delta\gamma_c(D_c)$ of the stationary loss rates of all the contents. In addition, Equation~(\ref{eqn: first derivative}) points out that these derivatives are proportional to $-\gamma_c$ in the limit of large replication / storage, and thus equalizing the loss rates should provide an approximate solution for the optimization problem. An immediate remark at this point is that it is unnecessary to store contents with very low popularity $\lambda_c$ if the loss rate of the other contents is already larger than $\lambda_c$.

An adaptive replication mechanism is characterized by two rules: a rule for creating new replicas and another one for evicting contents to make space for the new replicas. In order to figure out how to set these rules, we analyze the system in the fluid limit regime, with a separation of timescales such that the dynamics of the system with fixed replication have enough time to converge to their steady-state between every modification of the replication. Achieving this separation of timescales in practice would require slowing down enough the adaptation mechanism, which reduces the capacity of the system to react to changes. Therefore, we keep in mind such a separation of timescales as a justification for our theoretical analysis but we do not slow down our adaptive algorithms.

When trying to equalize the loss rates, it is natural to use the loss events to trigger the creation of new replicas. Then, new replicas for content $c$ are created at rate $\gamma_c$, and we let $\eta_c$ be the rate at which replicas of $c$ are deleted from the system. In the fluid limit regime under the separation of timescale assumption, the number of replicas of $c$ evolves according to $\dot D_c=\gamma_c-\eta_c$, where all the quantities refer to expectations in the steady-state of the system with fixed replication. At equilibrium, we have $\dot D_c=0$ and $\gamma_c=\eta_c$ for all the contents $c$, thus we need $\eta_c=\text{Cste}$ for all $c$ to equalize the loss rates. This would be achieved for example if we picked a content for eviction uniformly at random among all contents. However, the contents eligible for eviction are only those which are available (although some systems may allow modifying the caches of busy servers, as serving requests consumes upload bandwidth while updating caches requires only download bandwidth). Therefore, the most natural and practical eviction rule is to pick an \emph{available} content uniformly at random (hereafter, we refer to this policy simply as the RANDOM policy). Then, the eviction rate for content $c$ is given by $\eta_c\propto\pi(Z_c>0)=1-\frac{\gamma_c}{\lambda_c}$. So, at equilibrium, we can expect to have $\frac{\gamma_c}{1-\frac{\gamma_c}{\lambda_c}}=\text{Cste}$, $\forall c\in C$. In a large system, with large storage / replication and loss rates tending to zero, the difference with a replication trully equalizing the loss rates is negligeable. If we are confronted to a system with a large number of very unpopular contents though, we can compensate for this effect at the cost of maintaining additional counters for the number of evictions of each content.

Once the rule for creating replicas if fixed, we immediately obtain a family of adaptive algorithms by modifying the eviction rules from the cache network context as we did above for the RANDOM policy. Instead of focusing on ``usage'' and the incoming requests process as in cache networks with the LFU and LRU policies, we react here to the loss process. This yields the LFL (least frequently lost) and LRL (least recently lost) policies. RANDOM, LRL, and LFL are only three variants of a generic adaptive algorithm which performs the following actions at every loss for content $c$:
\begin{enumerate}
\item create an empty slot on an available server, using the eviction rule (RANDOM / LRL / LFL);
\item add a copy of the content $c$ into the empty slot.
\end{enumerate}
The three eviction rules considered here require a very elementary centralized coordinating entity. For the RANDOM rule, this coordinator simply checks which contents are available, picks one uniformly at random, say $c'$, and then chooses a random idle server $s$ storing $c'$. The server $s$ then picks a random memory slot and clears it, to store instead a copy of $c$. In the case of LRL, the coordinator needs in addition to maintain an ordered list of the least recently lost content (we call such a list an LRL list). Whenever a loss occurs for $c$, the coordinator picks the least recently lost available content $c'$ based on the LRL list (possibly restricting to contents less recently lost than $c$) and then updates the position of $c$ in the LRL list. It then picks a random idle server $s$ storing $c'$, which proceeds as for RANDOM. Finally, for LFL, the coordinator would need to maintain estimates of the loss rates of each content (by whatever means, e.g. exponentially weighted moving averages); when a loss happens, the coordinator picks the available content $c'$ with the smallest loss rate estimate and then proceeds as the other two rules. This last rule is more complicated as it involves a timescale adjustement for the estimation of the loss rates, therefore we will focus on the first two options in this paper. Note that the LFL policy does not suffer from the drawback that the eviction rate is biased towards popular contents due to their small unavailability, and this effect is also attenuated under the LRL policy. We point out that it is possible to operate the system in a fully distributed way (finding a random idle server first and then picking a content on it by whatever rule), but this approach is biased into selecting for eviction contents with a large number of available replicas (i.e. popular contents), which will lead to a replication with reduced efficency. It is of interest for future work to find a way to unbias such a distributed mechanism.

\subsection{Adapting Faster than Losses}\label{sec: virtual losses}

In the algorithms proposed in the previous subsection, we use the losses of the system as the only trigger for creating new replicas. It is convenient as these events are directly tied to the quantity we wish to control (the loss rates), however it also has drawbacks. Firstly, it implies that we want to generate a new replica for a content precisely when we have no available server for uploading it, and thus either we send two requests at a time to the data center instead of one (one for the user which we could not serve and one to generate a new copy) or we must delay the creation of the new replica and tolerate an inedaquate replication in-between, which also hinders the reactivity of the system. Secondly, unless in practice we intend to slow down the adaptation enough, there will always be oscillations in the replication. If losses happen almost as a Poisson process, then only a bit of slow-down is necessary to moderate the oscillations, but if they happen in batches (as it may very well be for the most popular contents) then we will create many replicas in a row for the same contents. If in addition we use the LRL or LFL rules for evicting replicas, then the same contents will suffer many evictions successively, fuelling again the oscillations in the replication. Finally, it is very likely that popularities are constantly changing. If the system is slow to react, then the replication may never be in phase with the current popularities but always lagging behind. For all these reasons, it is important to find a way to decouple the adaptation mechanisms from the losses in the system to some extent, and at least to find other ways to trigger adaptation.

We propose a solution, relying on the analysis in Section~\ref{sec: analysis by approximation}. A loss for content $c$ occurs when, upon arrival of a request for $c$, its number of available replicas is equal to $0$. In the same way, whenever a request arrives, we can use the current value of $Z_c$ to estimate the loss rate of $c$. Indeed, Equation~(\ref{eqn: approximation Z_c}) tells us how to relate the probability of $Z_c=z$ to the loss rate $\gamma_c$, for any $z\in\N$. Of course, successive samples of $Z_c$ are very correlated (note that losses may also be correlated though) and we must be careful not to be too confident in the estimate they provide. A simple way to use those estimates to improve the adaptation scheme is to generate \emph{virtual} loss events, to which any standard adaptive scheme such as those introduced in the previous subsection may then react. To that end, whenever a request for $c$ arrives, we generate a virtual loss with a certain probability $p_c(Z_c)$ depending on the current number of available replicas $Z_c$. The objective is to define $p_c(Z_c)$ so that the rates $(\widetilde\gamma_c)_{c\in C}$ of generation of virtual losses satisfy $\widetilde\gamma_c=\widetilde{\text{Cste}}\times\gamma_c$ for all $c\in C$ (so that the target replication still equalizes loss rates) and $\widetilde{\text{Cste}}$ is as high as possible (to get fast adaptation).

As a first step towards setting the probability $p_c(Z_c)$, we write Equation~(\ref{eqn: approximation Z_c}) as follows:
\begin{equation*}
\pi(Z_c=0)=\pi(Z_c=z)\prod_{i=1}^z\frac{\lambda_c+i\thetaeff}{D_c-i+1}.
\end{equation*}
This shows that, for any fixed value $z$ with $\pi(Z_c=z)\geq\pi(Z_c=0)$, we can generate events at rate $\gamma_c$ by subsampling at the time of a request arrival with $Z_c=z$ with a first probability
\begin{equation*}
q_c(z)=\prod_{i=1}^z\frac{\lambda_c+i\thetaeff}{D_c-i+1}.
\end{equation*}
If on the contrary $z$ is such that $\pi(Z_c=z)<\pi(Z_c=0)$, then the value of $q_c(z)$ given above is larger than $1$ as we cannot generate events at rate $\gamma_c$ by subsampling even more unlikely events. If we generated virtual losses at rate $\gamma_c$ as above for each value of $z$, then the total rate of virtual losses for content $c$ would be $\lambda_c\int_{z=1}^{D_c}\min\left\{\pi(Z_c=z),\pi(Z_c=0)\right\}$, which clearly still depends on $c$. We thus proceed in two additional steps towards setting $p_c(Z_c)$: we first restrict the range of admissible values of $Z_c$, for which we may generate virtual losses, by excluding the values $z$ such that $\pi(Z_c=z)<\pi(Z_c=0)$. In the regime $\thetaeff\geq1$, this can be done in a coarse way by letting $z^*_c=\frac{D_c-\lambda_c}{\thetaeff}$ and rejecting all the values $z>z^*_c$. Indeed,
\begin{equation*}
q_c(z^*_c)=\prod_{i=0}^{z^*_c}\frac{D_c-i\thetaeff}{D_c-i}\leq1,
\end{equation*}
and the distribution of $Z_c$ is unimodal with the mode at a smaller value $\frac{D_c-\lambda_c}{\thetaeff}\approx\rho z^*_c$. Now, subsampling with probability $q_c(Z_c)$ when a request arrives for content $c$ and $Z_c\leq z_c^*$ would generate events at a total rate $z^*_c\gamma_c$. Thus, it suffices to subsample again with probability $\frac{\min_{c'\in C}z^*_{c'}}{z_c^*}$ to obtain the same rate of virtual losses for all the contents (another approach would be to restrict again the range of admissible values of $z$, e.g. to values around the mode $\widehat z_c$).

To sum up, our approach is to generate a virtual loss for content $c$ at each arrival of a request for $c$ with probability
\begin{equation*}\label{eqn: virtual loss probability}
p_c(Z_c)=\frac{\min_{c'\in C}z_{c'}^*}{z_c^*}\ind(Z_c\leq z_c^*)q_c(Z_c).
\end{equation*}
The rate $\widetilde\gamma_c$ at which virtual losses are generated for content $c$ is then given by $\gamma_c\times\min_{c'\in C}z_{c'}^*$, which is independent of $c$ as planned. Whenever a virtual loss occurs, we can use whatever algorithm we wanted to use in the first place with real losses; there is no need to distinguish between the real losses and the virtual ones. For example, if we use the LRL policy, we update the position of $c$ in the LRL list and create a new replica for $c$ by evicting a least recently lost available content (from a server which does not already store $c$). If we choose to test for virtual losses at ends of service for $c$ (which yields the same outcome in distribution, as the system is reversible), the new replica can simply be uploaded by the server which just completed a service for $c$. Furthermore, in practice, we advocate estimating the values of $z_c^*$ and $p_c(Z_c)$ on the fly rather than learning these values for each content: $\thetaeff$ can be computed from $\rhoeff$, which is naturally approximated by the ratio of the current number of busy servers to the total number of servers; similarly, we can approximate $\lambda_c$ by the current number of requests for $c$ being served. From these, we can compute $z_c^*$ and $p_c(Z_c)$; it is only necessary to maintain an estimate for $\min_{c'\in C}z_{c'}^*$, which can be for example an average over the few least popular contents updated whenever a service ends for one of them.

In the next subsection, we evaluate the performance of the adaptative schemes proposed and the virtual losses mechanism.

\subsection{Evaluation of the Performance through Simulations}

Before getting to the simulation results, let us just mention that the complexity of simulating the adaptive algorithms grows very fast with the system size (with $n$, $m$ and $d$). Indeed, it is easy to see that simulating the system for a fixed duration $t$ requires $\Omega(md\overline\lambda t)$ operations. Furthermore, the time needed for the RANDOM algorithm to converge, when started at proportional replication, is roughly of the order of $\max_{c\in C}D_c/\overline\gamma$, where $\overline\gamma$ is the average loss rate for the limit replication, which decreases exponentially fast in $d$ as seen in Equation~(\ref{eqn: optimal gamma}). Therefore, if we want to compare all the adaptive schemes, we cannot simulate very large networks. Anyway, our intention is to show that our schemes work even for networks of reasonable size.

As in Section~\ref{sec: simulation approximation}, we used Zipf popularity distributions with exponents $0.8$ and $1.2$ and a class model to evaluate the performance of the various schemes. The results are qualitatively identical under all these models, so we only show the results for Zipf popularity with exponent $\alpha=0.8$. We compare the various schemes in terms of the replication achieved and its associated loss rates, as well as the speed at which the target replication is reached. We do not slow down the dynamics of the adaptive schemes even though this necessarily induces some oscillations in the replication obtained. Nonetheless, this setup is already sufficient to demonstrate the performance of the adaptive schemes. It would be interesting to quantify the potential improvement if one reduces the oscillations of the replication obtained (e.g. by quantifying the variance of the stationary distribution for the number of replicas for each content); we leave this out for future work. Also, we did not account for the load put on the data center to create new copies of the contents; one can simply double the loss rates for the adaptive schemes to capture this effect. Note that if adaptation speed is not a priority, one can trade it off to almost cancel this factor of $2$. Finally, concerning the virtual loss mechanism, we estimate all the quantities involved on the fly, as recommended in the previous section.

\begin{figure}
\centering
\[
\begin{array}{ccc}
\includegraphics[width=1.75in,keepaspectratio]{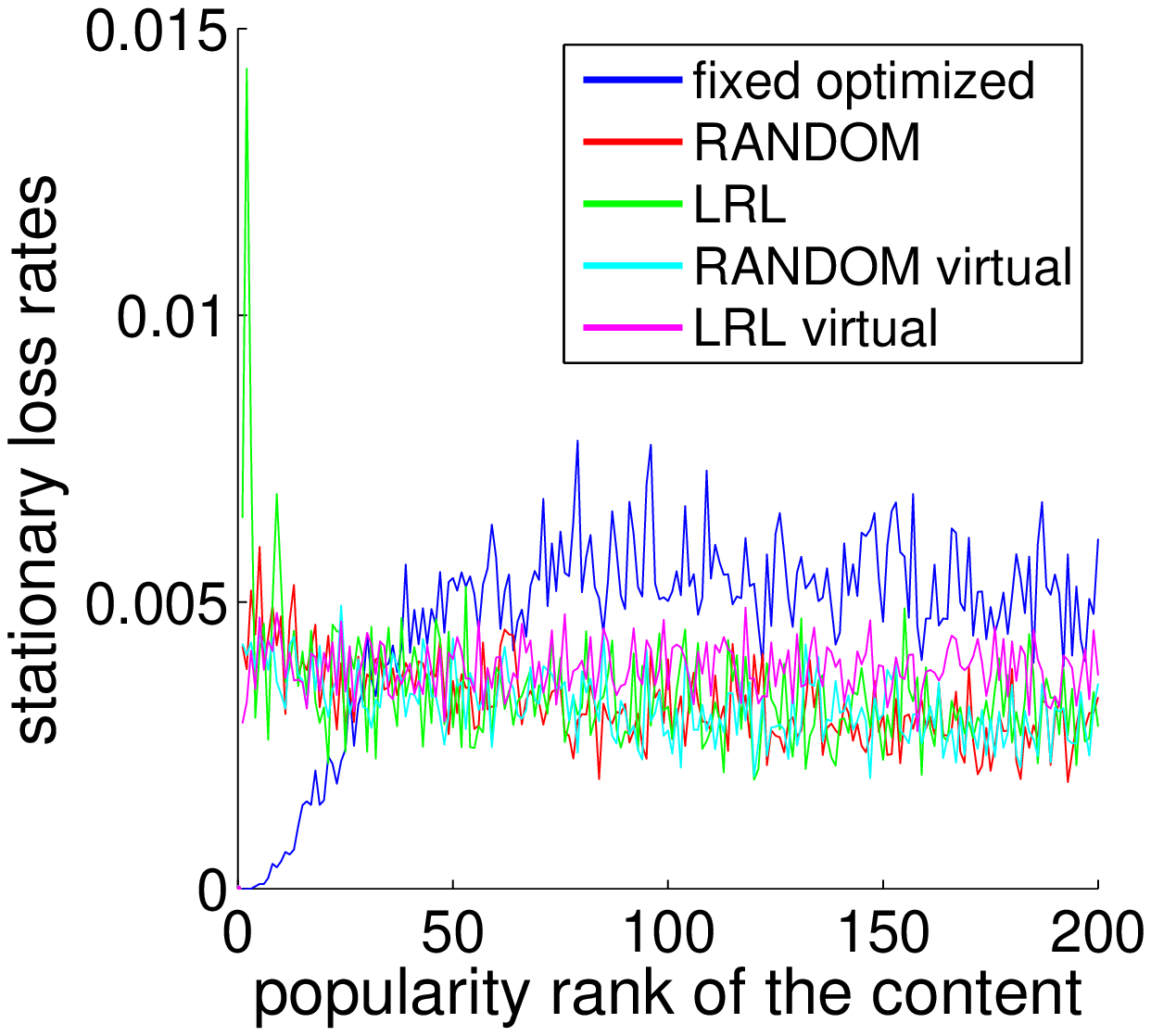}&
\includegraphics[width=1.55in,keepaspectratio]{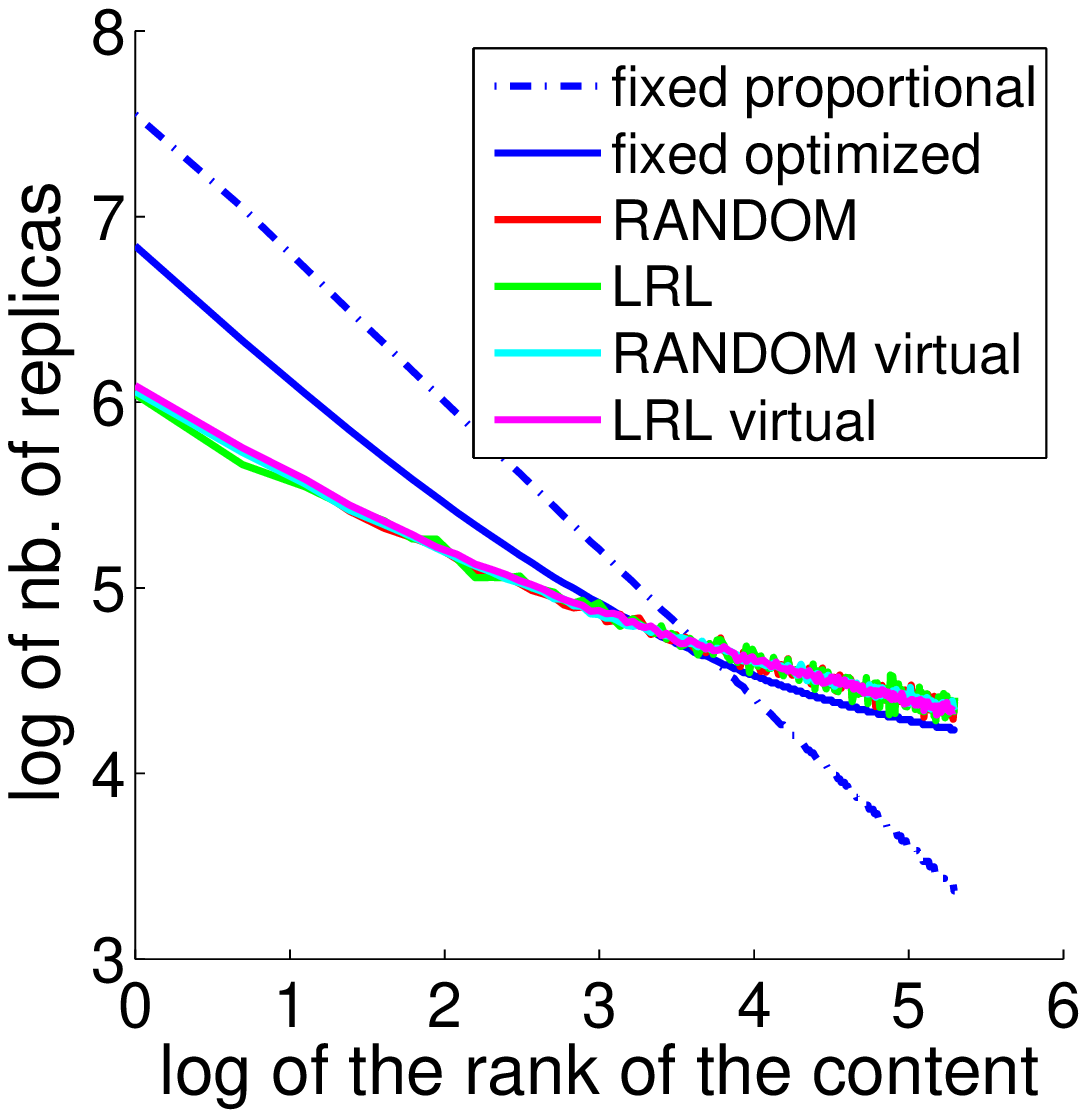}
\end{array}
\]
\caption{Adaptive schemes: loss rates and replication.}
\label{fig: loss rates and replication}
\end{figure}

%

\begin{figure}
\centering
\includegraphics[width=2.3in,keepaspectratio]{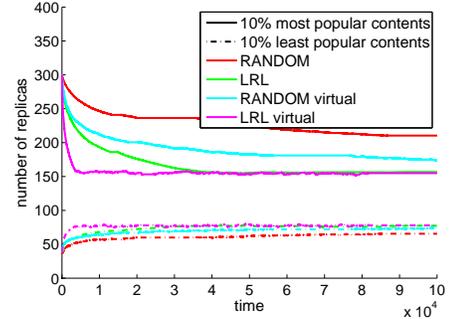}
\caption{Adaptive schemes: time evolution.}
\label{fig: time evolution}
\end{figure}


In Figure~\ref{fig: loss rates and replication}, we show results for the various adaptive schemes. On the left part of the figure, we show the stationary loss rates of all the contents; on the right part we show in log-log scale the stationary expectation of the numbers of replicas for each content. This plot shows firstly that all the adaptive schemes converge to the same replication and secondly that this replication equalizes the loss rates, as intended. In addition, the adaptive schemes perform even better than the optimized static replication, which suffers from finite network / finite storage effects, as they manage to find the right balance between popular and unpopular contents.

We compare the adaptation speed of the various schemes on Figure~\ref{fig: time evolution}, where we plot both the evolution of the average number of replicas of the $10\%$ most popular contents and that of the $10\%$ least popular ones, starting from proportional replication. As expected, the LRL schemes are faster than the RANDOM ones, but more importantly this plot clearly demonstrates the speed enhancement offered by the virtual loss method of Section~\ref{sec: virtual losses}. Regarding the benefits of such an enhanced reaction capability, there is an interesting property which we did not point out nor illustrate with the simulations: the virtual loss scheme has the potential to follow a constantly evolving popularity profile at no cost, as the required creations of replicas to adapt to the changing popularities can be done without requesting copies of the contents to the data center.

\section{Conclusion}

We addressed the problem of content replication in edge-assisted CDNs, with a special attention to capturing the most important constraints on server capabilities and matching policy. Based on large system and large storage asymptotics, we derived an accurate approximation for the performance of any given replication, thereby allowing offline optimization of the replication. In addition, levaraging the insights gathered in the analysis, we designed adaptive schemes converging to the optimal replication. Our basic adaptive algorithms react to losses, but we proposed further mechanisms to move away from losses and adapt even faster than they occur.


\begin{thebibliography}{10}
\providecommand{\url}[1]{#1}
\csname url@samestyle\endcsname
\providecommand{\newblock}{\relax}
\providecommand{\bibinfo}[2]{#2}
\providecommand{\BIBentrySTDinterwordspacing}{\spaceskip=0pt\relax}
\providecommand{\BIBentryALTinterwordstretchfactor}{4}
\providecommand{\BIBentryALTinterwordspacing}{\spaceskip=\fontdimen2\font plus
\BIBentryALTinterwordstretchfactor\fontdimen3\font minus
  \fontdimen4\font\relax}
\providecommand{\BIBforeignlanguage}[2]{{%
\expandafter\ifx\csname l@#1\endcsname\relax
\typeout{** WARNING: IEEEtran.bst: No hyphenation pattern has been}%
\typeout{** loaded for the language `#1'. Using the pattern for}%
\typeout{** the default language instead.}%
\else
\language=\csname l@#1\endcsname
\fi
#2}}
\providecommand{\BIBdecl}{\relax}
\BIBdecl

\bibitem{cisco}
\BIBentryALTinterwordspacing
{Cisco White Paper}, ``Cisco visual networking index: Forecast and methodology,
  2012-2017,'' May 2013. [Online]. Available:
  \url{http://www.cisco.com/en/US/solutions/collateral/ns341/ns525/ns537/ns705%
/ns827/white_paper_c11-481360.pdf}
\BIBentrySTDinterwordspacing

\bibitem{DilleyMPPSW02}
J.~Dilley, B.~Maggs, J.~Parikh, H.~Prokop, R.~Sitaraman, and B.~Weihl,
  ``Globally distributed content delivery,'' \emph{IEEE Internet Computing},
  vol.~6, no.~5, pp. 50--58, 2002.

\bibitem{TanM11}
B.~Tan and L.~Massouli{\'e}, ``Optimal content placement for peer-to-peer
  video-on-demand systems,'' in \emph{Proceedings IEEE INFOCOM}, 2011, pp.
  694--702.

\bibitem{LeconteLM12}
M.~Leconte, M.~Lelarge, and L.~Massouli{\'e}, ``Bipartite graph structures for
  efficient balancing of heterogeneous loads,'' \emph{ACM SIGMETRICS
  Performance Evaluation Review}, vol.~40, no.~1, pp. 41--52, 2012.

\bibitem{che2002hierarchical}
H.~Che, Y.~Tung, and Z.~Wang, ``Hierarchical web caching systems: Modeling,
  design and experimental results,'' \emph{IEEE Journal on Selected Areas in
  Communications}, vol.~20, no.~7, pp. 1305--1314, 2002.

\bibitem{fricker2012versatile}
C.~Fricker, P.~Robert, and J.~Roberts, ``A versatile and accurate approximation
  for lru cache performance,'' in \emph{Proceedings of the 24th International
  Teletraffic Congress}.

\bibitem{fricker2012impact}
C.~Fricker, P.~Robert, J.~Roberts, and N.~Sbihi, ``Impact of traffic mix on
  caching performance in a content-centric network,'' in \emph{IEEE Conference
  on Computer Communications Workshops (INFOCOM WKSHPS)}, 2012, pp. 310--315.

\bibitem{ZhouFC13}
Y.~Zhou, T.~Z. Fu, and D.~M. Chiu, ``On replication algorithm in p2p vod,''
  \emph{IEEE/ACM Transactions on Networking}, pp. 233 -- 243, 2013.

\bibitem{ZhouFC12}
------, ``A unifying model and analysis of p2p vod replication and
  scheduling,'' in \emph{Proceedings IEEE INFOCOM}, 2012, pp. 1530--1538.

\bibitem{CiulloMGLT12}
D.~Ciullo, V.~Martina, M.~Garetto, E.~Leonardi, and G.~L. Torrisi, ``Stochastic
  analysis of self-sustainability in peer-assisted vod systems,'' in
  \emph{Proceedings IEEE INFOCOM}, 2012, pp. 1539--1547.

\bibitem{jiang2012orchestrating}
W.~Jiang, S.~Ioannidis, L.~Massouli{\'e}, and F.~Picconi, ``Orchestrating
  massively distributed cdns,'' in \emph{Proceedings of the 8th international
  conference on Emerging networking experiments and technologies}.\hskip 1em
  plus 0.5em minus 0.4em\relax ACM, 2012, pp. 133--144.

\bibitem{rochman2013resource}
Y.~Rochman, H.~Levy, and E.~Brosh, ``Resource placement and assignment in
  distributed network topologies,'' in \emph{Proceedings IEEE INFOCOM}, 2013.

\bibitem{rochman2012max}
------, ``Max percentile replication for optimal performance in multi-regional
  p2p vod systems,'' in \emph{Ninth International Conference on Quantitative
  Evaluation of Systems (QEST)}.\hskip 1em plus 0.5em minus 0.4em\relax IEEE,
  2012, pp. 238--248.

\bibitem{laoutaris2005optimization}
N.~Laoutaris, V.~Zissimopoulos, and I.~Stavrakakis, ``On the optimization of
  storage capacity allocation for content distribution,'' \emph{Computer
  Networks}, vol.~47, no.~3, pp. 409--428, 2005.

\bibitem{cohen2002replication}
E.~Cohen and S.~Shenker, ``Replication strategies in unstructured peer-to-peer
  networks,'' in \emph{ACM SIGCOMM Computer Communication Review}, vol.~32,
  no.~4, 2002, pp. 177--190.

\bibitem{tewari2005fairness}
S.~Tewari and L.~Kleinrock, ``On fairness, optimal download performance and
  proportional replication in peer-to-peer networks,'' in \emph{NETWORKING
  2005. Networking Technologies, Services, and Protocols; Performance of
  Computer and Communication Networks; Mobile and Wireless Communications
  Systems}.\hskip 1em plus 0.5em minus 0.4em\relax Springer, 2005, pp.
  709--717.

\bibitem{tewari2006proportional}
------, ``Proportional replication in peer-to-peer networks.'' in
  \emph{Proceedings IEEE INFOCOM}, 2006.

\bibitem{LeconteLM13}
M.~Leconte, M.~Lelarge, and L.~Massouli{\'e}, ``Convergence of multivariate
  belief propagation, with applications to cuckoo hashing and load balancing,''
  in \emph{Proceedings of the Twenty-Fourth Annual ACM-SIAM Symposium on
  Discrete Algorithms (SODA)}, S.~Khanna, Ed.\hskip 1em plus 0.5em minus
  0.4em\relax SIAM, 2013, pp. 35--46.

\bibitem{kelly1991loss}
F.~P. Kelly, ``Loss networks,'' \emph{The annals of applied probability}, pp.
  319--378, 1991.

\bibitem{sznitman1991topics}
A.-S. Sznitman, ``Topics in propagation of chaos,'' in \emph{Ecole d'Et{\'e} de
  Probabilit{\'e}s de Saint-Flour XIX—1989}.\hskip 1em plus 0.5em minus
  0.4em\relax Springer, 1991, pp. 165--251.

\end{thebibliography}
\end{document}